\documentclass[10pt]{iopart}
\usepackage{amssymb}
\expandafter\let\csname equation*\endcsname\relax
\expandafter\let\csname endequation*\endcsname\relax
\usepackage{amsmath}
\usepackage{cancel}
\usepackage{braket}
\usepackage{graphicx}% Include figure files
\usepackage{dcolumn}% Align table columns on decimal point
\usepackage{bm}% bold math
\usepackage{braket}
\usepackage{color}
\usepackage{booktabs} % for much better looking tables
\usepackage{array} % for better arrays (eg matrices) in maths
\usepackage{paralist} % very flexible & customisable lists (eg. enumerate/itemize, etc.)
\usepackage{verbatim} % adds environment for commenting out blocks of text & for better verbatim
\usepackage{subfig} % make it possible to include more than one captioned figure/table in a single float
\usepackage{float} % to fix the conflict between hyperref and 
\usepackage{hyperref}
\hypersetup{colorlinks=true,linkcolor=blue,urlcolor=blue}

\begin{document}
\title[Space-averaged non-equilibrium Green's function approach for quantum transport in 3D]{Space-averaged non-equilibrium Green's function approach for quantum transport in 3D}
%---------------------
\author{Vahid Mosallanejad \textsuperscript{1,*}, Kuei-Lin Chiu \textsuperscript{2,*} and Wenjie Dou \textsuperscript{1,*}.}
%---------------------
\address{$^1$ School of Science, Westlake University, Hangzhou, Zhejiang 310024, China $\&$ Institute of Natural Sciences, Westlake Institute for Advanced Study, Hangzhou, Zhejiang 310024, China}
\address{$^2$ Department of Physics, National Sun Yat-Sen University, Kaohsiung 80424, Taiwan}
\ead{vahid@ustc.edu.cn, eins0728@gmail.com and douwenjie@westlake.edu.cn}
\vspace{10pt}
\begin{indented}
\item[] Apr 2025
\end{indented}

\date{\today}

\begin{abstract}
\textcolor{black}{The non-equilibrium Green's function (NEGF) approach offers} a practical framework for simulating various phenomena in mesoscopic systems. \textcolor{black}{As the dimension of electronic devices shrinks to just a few nanometers, the need for new effective-mass based 3D implementations of NEGF has become increasingly apparent}. This work extends our previous Finite-Volume implementation\textcolor{black}{---originally developed for the self-consistent solution of the Schr\"odinger and Poisson equations in 2D---into a full 3D NEGF framework}. Our implementation begins with exploring a few problems with the common textbook Finite Difference implementations of NEGF. We then concisely demonstrate how Finite-Volume discretization addresses few key \textcolor{black}{implementation} challenges. Importantly, we explain how this type of discretization enables evaluating the self-energies, which account for the effects of reservoirs. The potential applications of this new method are illustrated through \textcolor{black}{two} examples. \textcolor{black}{We anticipate that this implementation will be broadly applicable to open quantum systems, especially in cases where a fully three-dimensional domain is essential.}
\end{abstract}

%\keywords{Suggested keywords}%Use showkeys class option if keyword
                              %display desired

                              % Uncomment for keywords
\vspace{2pc}
\maketitle
\ioptwocol
%%%%%%%%%%%%%%%%%%%%%%%%%%%%%%%%%%
%\section{Introduction}
%\label{sec1}
%\section{Methods}
%\label{sec2} %
%In this section, we introduce FV-SP approach. Note that our approach is general, but we restrict ourselves to the 1DEG (a 2D problem) described below. 
%%%%%%%%%%%%%%%
%subsection{Theory of one-dimensional electron gas}
%\label{subsec21}
%%%%%%%%%%%%%
%\subsection{Cell-centered discretization scheme in 2D}

\section{\label{sec:level1}Introduction}
\textcolor{black}{The non-equilibrium Green’s function (NEGF) formalism is widely used in solid-state physics and chemistry, however, it does not always correspond to a single, well-defined mathematical procedure}~\cite{odashima2016pedagogical,peskin2010introduction, perfetto2015first,cohen2020green,mosallanejad2024floquet}. 
\textcolor{black}{Its versatility makes it applicable to a broad range of systems, including} ultra-small transistors~\cite{luisier2014atomistic, ahn2025nonequilibrium}, spintronics~\cite{chen2009spin,vadde2023orthogonal}, thermoelectric materials~\cite{neophytou2016modulation,foster2019effectiveness,polanco2021nonequilibrium}, and molecular electronics~\cite{evers2004conductance, sand2021multiconfiguration}, 2D materials~\cite{shah2019design, zhang2023electronic}, disordered systems~\cite{zhou2016theory, shah2019transport, zhang2024random}, and optoelectronic devices~\cite{aeberhard2011theory}.
%-----------------
%----------------
The Green functions should be defined according to the methodology and specific aspects of the system, such as the model Hamiltonian. 
%----------------------------------
\textcolor{black}{The one-body NEGF method gained popularity in mesoscopic systems in the late 20th century, originating from the pioneering work of Caroli in the 1970s~\cite{caroli1971direct} and later advanced by Datta, Meir, Wingreen, and Jauho ~\cite{datta1992exclusion, wingreen1993time, jauho1994time} }. 
%------------------------
%-----------------------
Here, we refer specifically to the space and energy-resolved NEGF for mesoscopic systems~\cite{datta2005quantum}.
%-----------------------
The more fundamental (two-time or contour-ordered) NEGF formalism is rooted in many-body perturbation theory, established through the works of Martin, Schwinger, Keldysh, Kadanoff, Baym and others \cite{martin1959theory, keldysh2024diagram, kadanoff2018quantum}. 
%----------------------
\textcolor{black}{The procedure for obtaining the energy-resolved NEGF---for example, from the Kadanoff-Baym formalism---involves performing a Fourier transform of the two-time Green’s functions~\cite{pourfath2014non, ryndyk2016theory}.}
% The procedure to obtain the energy-resolved NEGF (for instance, from the Kadanoff-Baym NEGF) involves Fourier transforming the two-time Green’s functions~\cite{pourfath2014non, ryndyk2016theory}.
%----------------------------------------
The space-energy-resolved NEGF primarily addresses steady-state quantum transport properties, such as transmission probabilities, which are essential for interpreting transport in low-dimensional devices (e.g., molecular junctions). A key strength of NEGF lies in its treatment of open quantum systems via the concept of \textit{contact self-energy}. NEGF is widely regarded as one of the most powerful and accurate methods \textcolor{black}{in quantum transport} and finds broad applications in both scientific and industrial contexts ~\cite{shao2025scaled, li2015towards, pal2024three}.
%------------------------
The NEGF technique became computationally expensive in higher dimensions to the point that additional mathematical tricks such as the coupled mode-space approach~\cite{venugopal2002simulating,luisier2006quantum} are required to practically perform a quantum transport study in 3D systems \textcolor{black}{like quantum wires}. 
%-------------------------
\textcolor{black}{Moreover, implementing NEGF becomes increasingly cumbersome when material properties---such as effective mass, electron affinity, or others---vary across the device geometry}
%----------
%Moreover, NEGF becomes cumbersome to implement when the effective mass, electron affinity, or other material properties vary across the geometry~\cite{martinez2007study}.
%-----------------------
In realistic systems, ideal ballistic transport is disrupted by local imperfections (lattice disorder, dopant inhomogeneity, or interfacial strain). To accurately model these effects in large 3D systems, a robust NEGF formalism must incorporate spatially resolved material properties, allowing for the self-adjustment of local properties.
%------------
\textcolor{black}{It is the intention of this letter to introduce the concept of space-averaged Green's function associated with the cell-centered Finite-Volume (FV) discretization method.} 
%------------------
\textcolor{black}{Although a few publications have reported the existence of the Voronoi FV-NEGF approach~\cite{baumgartner2013vsp}, it has not seen further development over the past decade. For example, in one study, the authors used the finite-volume method to solve the Poisson equation, but not the transport equation~\cite{berrada2020nano}. } 
%-----------------
%Although a limited number of publications have reported the existence of Voronoi FV NEGF \cite{baumgartner2013vsp} but this approach does not received any further developments within a decade. For example, in one work, the authors have used FV to solve the Poisson equation but not the transport equation~\cite{berrada2020nano}. 
%--------------------------
We \textcolor{black}{propose} that the averaged Green's function formalism can improve the implementation procedure of the effective mass NEGF in \textcolor{black}{a 3D domain}. The proposed method can be regarded as an extension of our previous work in solving Schr\"odinger-Poisson systems in  \textcolor{black}{a 2D domain}~\cite{mosallanejad2023cell}. 
We demonstrate that this approach can \textcolor{black}{overcome} several limitations of the Finite-Difference (FD) NEGF method. While the current work does not explore \textcolor{black} {the full} coupled self-consistent Poisson-NEGF simulations for conciseness, we argue that the FV scheme can fully realize its potential in realistic Poisson-NEGF applications. Note that the existing Finite Element (FE) NEGF requires using the concept of shape functions, which makes the implementation more complex or less intuitive ~\cite{polizzi2003multidimensional,lee2025new}. 

%%%%%%%%%%%%%%%%%%%%%%%%%%%%%%%%%%%%%%%%%%%%%%%%%
%%%%%%%%%%%%%%%%%%%%%%%%%%%%%%%%%%%%%%%%%%%%%%%%%

\section{\label{sec:level2}Method}
\subsection{\label{sec:level2} Equation of motion for NEGF: problems with the FD NEGF}
In the context of mesoscopic quantum transport, the retarded Green's function, $G^R$, is \textcolor{black}{often} considered as the most basic Green's function and thus \textcolor{black}{plays} the central role. 
%--------------------
A straightforward way \textcolor{black}{to derive} the equation of motion for the retarded Green's function in a mesoscopic system is to begin with the time-independent Schr\"odinger equation, using the Hamiltonian operator, $\hat{H}$, for an infinite physical domain.
%--------------------
Then, we turn the wave (envelope) function into the retarded Green's function ($\hat{\psi}\!\rightarrow\!\hat{G}$). \textcolor{black}{Simultaneously,} we add a positive imaginary one-site energy (denoted below by $i\eta$) to the Hamiltonian and add the Dirac delta function (as the source of impulse) to the right-hand side of the equation. 
%--------------------
This easy justification does not include the particle-particle interactions and only \textcolor{black}{takes account of} particle exchanges between the contacts and the reduced quantum system. 
%----------------------
\textcolor{black}{Particle interactions can later be incorporated into this version of NEGF as phenomenological models, provided that conservation laws are carefully respected.}
%----------------------
The equation of motion for $\hat{G}^R$ reads as:
\begin{eqnarray}
\label{eq:1}
\big( (E+i \eta)\hat{S}-\hat{H} \big) 
\hat{G}^R(E; \mathbf{\hat{r}},\mathbf{\hat{r}}^{\prime})=\delta (\mathbf{\hat{r}}-\mathbf{\hat{r}}^{\prime}),
\end{eqnarray}
where $\hat{S}$ is the overlap operator and the energy, $E$, treated as a continuous variable. Hereafter, we drop the superscript $^R$ and the implicit dependence on the energy variable $E$ in the notation for the retarded Green's function, ${G}^R(E;\mathbf{r},\mathbf{r}^{\prime})\!\equiv\! {G}(\mathbf{r},\mathbf{r}^{\prime})$, simplifying the presentation of the subsequent relations.
%------------------------
Then, one should act \textcolor{black}{$\langle\mathbf{r}|$} and $|\mathbf{r}^{\prime}\rangle$ from left and right such that $\hat{H}\!\rightarrow\!H$, $\hat{S}\!\rightarrow\!S$ and $\hat{G}\!\rightarrow\!G$. \textcolor{black}{This is equivalent to defining a model Hamiltonian in terms of a chosen set of orbitals (basis set).}
%---------------------------
For example, in the tight-binding (TB) model, $H$ and $S$ are known matrices defined by on-site orbital energies and hopping terms between orbitals, where a set of orbitals forms the basis. Within the TB framework, it is reasonable to replace $\langle\mathbf{r}|\delta (\mathbf{\hat{r}}-\mathbf{\hat{r}}^{\prime})|\mathbf{r}^{\prime} \rangle$ with the identity matrix $I$. 
%-------------
\textcolor{black}{The procedure involves partitioning the physical system and applying matrix algebra to reduce the formally infinite equation of motion to a finite-size equation that describes the reduced quantum system.}
%----------------
%The procedure follows by \textit{partitioning} the physical system, using matrix algebra to reduce the formally infinite equation of motion to a finite-size equation that describes the reduced quantum system. 
%----------------------------------------
For a simple one-band effective mass Hamiltonian, one needs to identify a parabolic-type Hamiltonian in terms of the effective mass, $\hat{S}\!\rightarrow \!I$, and discretize Eq.~(\ref{eq:1}) in an infinite domain. 
%---------
To be precise, it is customary to use the Hamiltonian given by: $H(\mathbf{r})\!=\!(\hbar^2/2m^{*}(\mathbf{r}))\nabla^2\!+\!U(\mathbf{r})$, where $U(\mathbf{r})\!=\!-eV(\mathbf{r})$ represents the mean-field potential energy. Then, the Laplacian operator is often approximated using a FD scheme. 
%---------------------
For example, \textcolor{black}{in} a 1D domain, the Laplacian operator can be expressed as: \textcolor{black}{$\nabla^2u \!\approx\! (u_{i+1}\!+\!2u_{i}\!-\!u_{i-1})/ \Delta x^2$}. \textcolor{black}{Next, Dirac delta function is replaced by an identity matrix}. By discretizing the domain, partitioning it into semi-infinite contacts and a reduced system, and applying matrix algebra, we can arrive at the limited size matrix relation: $([A]\!-\![\Sigma])[G]\!=\!\mathbf{1}$. Here, $\Sigma$ represents the total \textit{self-energy} of the contacts, and $[A]\!=\!(E+i\eta)[I]\!-\![H_C]$, where the subscript $_C$ denotes the reduced system (channel/center).
%------------------------
While the above procedure works in practice, it overlooks the conceptual nature of the Dirac delta \textcolor{black}{function}. Specifically, the Dirac delta \textcolor{black}{function} is not a mathematical function in the traditional sense to be replaced by an approximation. In particular, Dirac delta \textcolor{black}{function} must satisfy the property: $\int \delta (\mathbf{r}\!-\!\mathbf{r}^{\prime})d \mathbf{r}\!=\!1$. 

This seemingly unimportant issue with the Dirac delta leads to a unit problem in evaluating electron density based on the relation $n(\mathbf{r})\!=\!(1/2\pi)\int \operatorname{Tr}\big(\!-\!iG^{<}(\mathbf{r},\mathbf{r})\big)dE$. That is, in the conventional FD-NEGF method, the scaling factor $\hbar^2/2m_e\Delta x^2$ ($m_e$ is the bare electron mass) has units of energy, so the Green's functions would only have units of inverse energy. However, the correct units should be inverse energy multiplied by inverse length (or inverse volume in 3D). For these reasons, directly approximating the Dirac delta \textcolor{black}{function} itself is not entirely appropriate. A justification for this shortcoming in a 1D discrete mesh may be provided by the following definition of the Dirac delta:
\begin{eqnarray}
\label{eq:2}
\delta(x-x^{\prime})=\lim _{\underline{\Delta} x_i \rightarrow 0^{+}} 
\begin{cases}
\frac{1}{ \underline{\Delta} x_i} & \text {if~~} x_i=x_j^{\prime} 
\\ 
0                                 & \text {otherwise }
\end{cases}.~~
\end{eqnarray}
Here, $\underline{\Delta} x_i\!=\!x_{i+1/2}-x_{i-1/2}$ refers to the distance between midpoints. \textcolor{black}{In a 3D geometry, we must extract all the coordination of midpoints to obtain the volume around each grid, $\underline{\Delta} V_i$. Then the correct Green's function obtains by $[A\!-\!\Sigma]^{-1}[\underline{\Delta} V]$, where $[\underline{\Delta} V]$ refers to a diagonal matrix made of these volumes.} Nonetheless, such \textcolor{black}{redemption} make the correct implementation cumbersome as it requires involvement to the midpoint grids. 
%-----------------------------

Moreover, when modeling a complex electronic device where multiple domains represent different materials, the most convenient meshing approach is to mesh each domain independently. This ensures that discrete points are positioned exactly at the material interfaces. However, with the FD scheme, it is unclear how to handle discontinuities in material properties. Specifically, what effective mass should be used when we approximate $(\hbar^2/2m^*(\mathbf{r}))\nabla^2$ at the interface between two materials? In contrast, as will be clarified shortly, the FV method provides a clear and well-defined protocol for addressing such issues. Here, it is important to emphasize that discontinuities or abrupt changes in material properties at the interface are the origin of quantum confinement effects and must be treated carefully.

%%%%%%%%%%%%%%%%%%%%%%%%%%%%%%%%%%%%%%%%%%%%
\subsection{\label{sec:level2} Finite-Volume implementation of NEGF}
\begin{figure}[h]
	\begin{center}
		\includegraphics[width=7.5cm]{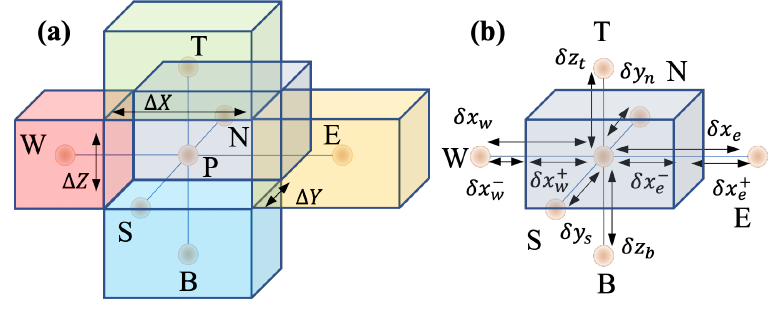}
	\end{center}
	\caption{\label{fig1} (a) Central cell, $P$, and five neighbour FV cells. While the southern cell's center is depicted by small sphere but the southern cell itself does not plotted. (b) relative distances between $P$ and other neighbours.}
\end{figure}

The correct solution to this implementation problem can be found by using the cell-centered FV discretization scheme. To begin, the more appropriate single band effective mass Hamiltonian can be \textcolor{black}{written as:}
\begin{eqnarray}
\label{eq:3}
H(\mathbf{r})=-\frac{\hbar^2}{2} \nabla \cdot \big(\frac{1}{m^*(\mathbf{r})} \nabla\big)+U(\mathbf{r}).
\end{eqnarray}
Fortunately, above form allows us to employ the Divergence theorem \textcolor{black}{which} will be explained shortly. In addition, this \textcolor{black}{formula} preserves continuity of the current at the interface between two different materials~\cite{burt1992justification}. The first step of FV method is to take the integral from both sides of Eq.~(\ref{eq:1}) \textcolor{black}{over} a central cell, depicted as the cell $P$ in Fig. \ref{fig1} (a). The cell $P$, with volume ${\Delta V_P}\!=\!\Delta x \Delta y \Delta z$, is called the \textit{control volume}, in computational fluid dynamics. 
%--------------------------
After taking the integration, the left and right hand sides (LHS, RHS) of Eq.~(\ref{eq:1}) can respectively be denoted as
\begin{eqnarray}
\label{eq:4_5_1}
&\int_{V_P}
\big( 
(E+i\eta)I+
\nabla \cdot (\mathbf{\Gamma}^*(\mathbf{r}) \nabla)- U(\mathbf{r}) 
\big) 
G(\mathbf{r},\mathbf{r}^{\prime})~dV,~~\\ 
\label{eq:4_5_2}
&\int_{V_P} \delta (\mathbf{r}-\mathbf{r}^{\prime})~dV,
\end{eqnarray}
%--------------------------GGGGGGGG
%---------------------------
where $\Gamma^*(\mathbf{r})\!=\!\hbar^2/2m^*(\mathbf{r})$. Note that, $\mathbf{\Gamma}^*(\mathbf{r})$ refers to a none-isotropic quantity associated with non-isotropic effective mass. We then desire to reduce the $G(\mathbf{r},\mathbf{r}^{\prime})$, as a continuous function of $\mathbf{r}$ and $\mathbf{r}^{\prime}$, to a matrix value associated with $(P,P^{\prime})$ as we run the discreet equation for all control volumes in the infinite domain. 
%--------------------------
With that, the definition of the RHS reduces correctly to the identity matrix on discreet spaces of $P$ and $P^{\prime}$. 
\begin{eqnarray}
\label{eq:6}
\int_{V_P} \delta (\mathbf{r}-\mathbf{r}^{\prime})dV
=
\begin{cases} 
1 & \text{if}~P=P^{\prime} \\
0 & \text{otherwise}
\end{cases}
.
\end{eqnarray}
The potential energy term is approximated by the following \textcolor{black}{piecewise} approximation
\begin{eqnarray}
\label{eq:7}
\int_{V_P} U(\mathbf{r})
G(\mathbf{r},\mathbf{r}^{\prime}) dV
\approx
\overline{U}(P)
\overline{G}(P, P^{\prime}) 
\Delta V_P,
\end{eqnarray}
which means that we have associated averages of Retarded Green's function, $\overline{G}$, and the potential energy, $\overline{U}(P)$, with their values at the center of the cell $P$. Same \textcolor{black}{piecewise} approximation will be used for the following diagonal terms
\begin{eqnarray}
\label{eq:8}
\int_{V_P} \!(E+i\eta)I~
G(\mathbf{r},\mathbf{r}^{\prime}) dV
\!\approx\!
(E+i\eta)I~
\overline{G}(P, P^{\prime}) 
\Delta V_P.~~~
\end{eqnarray}
%-------------------------
% The integral of kinetic energy term simplifies based on Divergence theorem as
\textcolor{black}{The integral of the kinetic energy term is simplified using the divergence theorem, resulting in:}
\begin{eqnarray}
\label{eq:9}
\iiint_{V_P}\nabla \cdot (\mathbf{\Gamma}^*(\mathbf{r}) \nabla){G} \mathrm{d} V=
\iint_{S_P}\mathbf{\Gamma}^*(\mathbf{r}) \nabla{G} \cdot \hat{\mathbf{n}}\mathrm{d}S, ~~~
\end{eqnarray}
in which $S_P$ refers to all six faces of the cell $P$ and $\hat{\mathbf{n}}$ is the normal vector to each of these faces. What remains is to approximate the RHS of Eq.~(\ref{eq:9}) in terms of the value of ${G}$ at the centers of \textcolor{black}{closest} cells. 
%--------------
\textcolor{black}{At this step, one must be cautious about keeping the continuity of flux at interfaces between cells.}
%--------------
%At this step, one must be cautious about maintaining flux continuity across cell interfaces.
%-------------
Within the cell-centered FV method, that requirement \textcolor{black}{is enforced} by evaluating the values of $\mathbf{\Gamma}^* \!\equiv\! \{ \Gamma^*_x, \Gamma^*_y, \Gamma^*_z \}$ at the six interfaces via the \textit{harmonic mean} approximation.
%-------------
To be specific, $\Gamma^*_x$ at the eastern interface \textcolor{black}{is given} by
\begin{eqnarray}
\label{eq:10}
\Gamma^*_{x_e}= \frac{\Gamma_E^* \Gamma^*_P}{\beta \Gamma^*_E+(1-\beta) \Gamma^*_P},
\end{eqnarray}
where $\beta\!=\!\delta x_{e^{-}} / \delta x_e$ and $1-\beta\!=\!\delta x_{e^{+}} / \delta x_e$, see geometrical distances in Fig. \ref{fig1} (b).
%--------------
$\Gamma^*_x$ at the western interface, $\Gamma^*_{x_w}$, \textcolor{black}{is evaluated} by the same relation except that $\Gamma^*_E\!\rightarrow\! \Gamma^*_W$, 
$\beta\!=\!\delta x_{w^{+}} / \delta x_w$, 
and 
$1\!-\!\beta\!=\!\delta x_{w^{-}} / \delta x_w$. 
%---------------------------------------
$\Gamma^*_y$ at the southern and northern interfaces, and $\Gamma^*_z$ at the top and bottom interfaces must be evaluated in a same way. The details regarding flux continuity at the interfaces can be found in Ref.~\cite{mosallanejad2023cell}.
%---------------------------------------
A key advantage of FV method is that evaluating $\mathbf{\Gamma}^*$ at the interfaces, using values from the centers of adjacent cells, resolves a major limitation of the FD discretization scheme, as discussed earlier.
%------------------------------------ Here, t
Then, RHS of Eq.~(\ref{eq:9}) can be approximated by
\begin{eqnarray}
\label{eq:11}
\begin{aligned}
&\big(
\Gamma_{x_e}\!\!\frac{\overline{G}_E\!-\!\overline{G}_P}{\delta x_e}
\!-\!
\Gamma_{x_w}\!\!\frac{\overline{G}_P\!-\!\overline{G}_W}{\delta x_w}
\big) 
\Delta A_{yz}\!+\! 
\big(
\Gamma_{y_n}\!\!\frac{\overline{G}_N\!-\!\overline{G}_P}{\delta y_n} -\!
\\
&
\Gamma_{y_s}\!\!\frac{\overline{G}_P\!-\!\overline{G}_S}{\delta y_s}
\big)
\Delta A_{xz}\!+\!
\big(
\Gamma_{z_t}\!\!\frac{\overline{G}_T\!-\!\overline{G}_P}{\delta z_t}
\!-\!
\Gamma_{z_b}\!\!\frac{\overline{G}_P\!-\!\overline{G}_B}{\delta z_b}
\big)
\Delta A_{xy},
\end{aligned}
\end{eqnarray}
%-------------------
where $A_{yz}\!=\!\Delta y\Delta z$, and $A_{xz}$ and $A_{xy}$ represent the appropriate areas. \textcolor{black}{After substituting approximations given in Eqs.~(\ref{eq:6})-(\ref{eq:8}) and Eq.~(\ref{eq:11}) into Eqs.~(\ref{eq:4_5_1}) and (\ref{eq:4_5_2}), equating them, and dividing all terms by $\Delta V_P$, Eq.~(\ref{eq:1}) is simplified to} 
\begin{eqnarray}
\label{eq:12}
\begin{aligned}
&-a_W\overline{G}_W-a_B\overline{G}_B-a_S\overline{G}_S \!+\! \bigl( (E\!+\!i\eta)+ a_P \!-\!\overline{U}_P \bigr) \overline{G}_{P} 
\\&
-a_N\overline{G}_N -a_T\overline{G}_T-a_E \overline{G}_E=\Delta V_P^{-1}, 
\\&
a_P=a_W+a_B+a_S+a_N+a_T+a_E,
\end{aligned}
\end{eqnarray}
where material and geometrical coefficients are combined into a series of coefficients associated with six neighbor cells given by 
\begin{eqnarray}
\label{eq:13}
\begin{aligned}
a_{W,E}
\!=\!\!
\frac{\Gamma_{x_{w,e}} }
{\delta x_{w,e} \Delta X}, 
a_{S,N}
\!=\!\!
\frac{\Gamma_{y_{s,n}} }
{\delta y_{s,n} \Delta Y}, 
a_{B,T} 
\!=\!\!
\frac{\Gamma_{z_{b,t}} }
{\delta z_{b,t} \Delta Z}.~~~~~
\end{aligned}
\end{eqnarray}
%---------------
\textcolor{black}{Eq.~(\ref{eq:12}) is interesting because $\Delta V_P^{-1}$ appears on the RHS. This means the Green's function represents the space-averaged variable and has the correct units for evaluating electron density. We refer to these coefficients as the \textit{a-coefficient}s and they form six vectors as we run over all cells. It is worth noting the analogy between the a-coefficients and tight-binding parameters: the central coefficient, $a_{P}$, corresponds to the \textit{on-site}, while the other a-coefficients represent the \textit{hopping} energies. }
%--------------------
% Note that, there is analogy between a-coefficients and TB parameters where the central coefficient denoted by the $a_{P}$ and other a-coefficients play the role of the \textit{on-site} and \textit{hopping} energies, respectively.  
%--------------------
Eq.~(\ref{eq:12}) must run over all cells, and eventually leads to a matrix equation $[A][\overline{G}]\!=\![\Delta V]^{-1}$ with the formally infinite 7-diagonal matrix $[A]$. Hereafter, we refer to this implementation as Finite-Volume NEGF (FV-NEGF).  
%------------------------
\subsection{\label{sec:level2} \textcolor{black}{Applying boundary conditions and reducing FV-NEGF to the transport domain}}
\begin{figure}[h]
	\begin{center}
		\includegraphics[width=6.0cm]{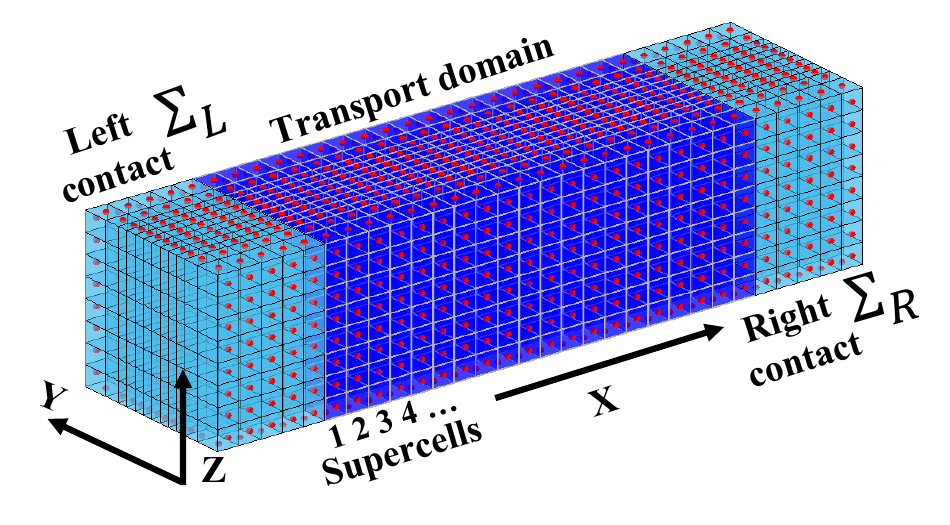}
	\end{center}
	\caption{\label{fig2} Schematics of a 3D domain resembling a nanowire which is discretized into nonuniform rectangular control volumes. Outermost cells on western and eastern faces must be subjected to the open boundary condition while the Dirichlet boundary condition must be imposed to the rest of outermost cells indicated by red dots.}
\end{figure}

%------------------------
Fig.~\ref{fig2} shows a schematic picture of a nanowire where we use it as an instrumental tool to illustrate the implementation of different boundary conditions. \textcolor{black}{The south, north, top, and bottom faces must be treated with the zero Dirichlet boundary condition, while the west and east faces must be treated with the open boundary condition.}
%----------------
\textcolor{black}{Before reducing $[A]$ to a finite-sized matrix for the reduced domain, we need to impose the Dirichlet boundary condition on the non-open boundary faces.}
%----------------
%---------------------------
%-------------

\textcolor{black}{ 
The zero Dirichlet boundary condition is applied to Eq.~(\ref{eq:12}) in two steps: (I) by eliminating any term in the set $\{a_{S},a_{N},a_{B},a_{T}\}$ relevant to the faces with Dirichlet boundary condition, from the off-diagonal, (II) by modifying the same term within the diagonal term ($a_P$) by $a_{S,N}\!=\!\Gamma_{s,n}/({\delta y^{+,-}_{s,n}}\Delta Y)$ and $a_{B,T}\!=\!\Gamma_{b,t}/({\delta z^{+,-}_{b,t}}\Delta Z)$.}
%---------------------------
Here, the values for $\mathbf{\Gamma}^*$ at Dirichlet boundaries ($\Gamma_{s,n,b,t~}$) are known, and therefore there is no need to evaluate them by the harmonic mean. \textcolor{black}{Details regarding the} implementation of Dirichlet boundary condition can be found in Ref.~\cite{mosallanejad2023cell}.
%---------------------------
Thus, \textcolor{black}{handling} the closed boundary condition essentially reduces to correcting the six vectors of a-coefficients. We stress that the above equation is still for the semi-infinite domain. 
%-------------------------
\textcolor{black}{To handle the open boundary conditions on the western and eastern faces, the total domain should be partitioned into semi-infinite left and right contacts, along with the reduced transport domain (i.e., the channel).}
%------------
Then the infinite matrix equation $[A][\overline{G}]\!=\![\Delta V]^{-1}$ \textcolor{black}{can} be rewritten as the following partitioned equation: 
%-------------------------
\setlength\arraycolsep{0.5pt}% some length
\begin{eqnarray}
\label{eq:14}
  \left[\!\begin{smallmatrix}
   A_{L }    & A_{L C} & O \\
   A_{C L} & A_{C}    & A_{C R} \\
   O          & A_{R C}  & A_{R}
  \end{smallmatrix}\!\right]
  \!\!
  \left[\!\begin{smallmatrix}
    \overline{G}_{L}    & \overline{G}_{LC}  & \overline{G}_{CR}\\
    \overline{G}_{C L} & \overline{G}_{C}   & \overline{G}_{CR}\\
    \overline{G}_{R L} & \overline{G}_{RC} & \overline{G}_{L}
  \end{smallmatrix}\!\right]
\!\!=\!\!
  \left[\!\begin{smallmatrix}
   \Delta V_{L} & O & O \\
   O & \Delta V_{C} & O \\
   O & O & \Delta V_{R}
  \end{smallmatrix}\!\right]^{\!-1}\!\!.~~~~~
\end{eqnarray}
Focusing on the central column of $\overline{G}$, we can arrive at three equations for $\overline{G}_{LC}$, $\overline{G}_{C}$, and $\overline{G}_{RC}$. 
%%%%%%%%%%%%%%%%%%%%%%%%%%%%
Using matrix algebra the first and third matrix equations can be combined into the second equation as 
\begin{eqnarray}
\label{eq:15}
\left[A_C\!-\!A_{C L} A_L^{-1} A_{L C}\!-\!A_{C R} A_R^{-1} A_{R C}\right] \overline{G}_C\!=\!\left[\Delta V_{C} \right]^{-1}\!.~~~~~
\end{eqnarray}
It is conventional to define the left and right \textit{self-energies} as $\Sigma_L\!=\!A_{C L} A_L^{-1} A_{L C}$ and $\Sigma_R\!=\!A_{C R} A_R^{-1} A_{R C}$\textcolor{black}{, respectively,} such that above relation can be written as 
\begin{eqnarray}
\label{eq:16}
[A_C-\Sigma_L -\Sigma_R][\overline{G}_C][\Delta V_{C}]=[1].
\end{eqnarray}
%----------------------------
\textcolor{black}{We refer to} $A_{L,R}^{-1}$ as the contact Green's function. 
%----------------------------
\textcolor{black}{Note that $\Delta V_{L}$ and $\Delta V_{R}$ did not appear in Eq.~(\ref{eq:16})}. 
%----------------------------
\textcolor{black}{Computing the full contact Green's function is intractable as the contact Hamiltonian can be very large.}
%%%%%%%%%%%%%%%%%%%%%%%%%%%%%

We can further divide the transport domain into a set of supercells, marked by integers in Fig.~\ref{fig2}. Here, a supercell is defined as \textcolor{black}{the collection of all control volumes sharing the same x-coordinate, $x_P$.}
%----------------------------
In the same way, we can divide the contact domains into a set of supercells, although \textcolor{black}{they are} not shown in Fig. \ref{fig2}.
%---------
This extra division allows us to make full use of the concept of the contact \textit{surface} Green's function, $g_{L,R}$, in evaluating $\Sigma_{L,R}$. 
%----------------
Surface Green's function essentially imply that one only needs a few blocks of the contact Green's functions to evaluate the self-energy matrices. Then, the equation for the non-vanishing blocks of the self-energies can be expressed as: $[\Sigma_L]_{11}\!=\!A_{1,0}~g_{L}~A_{0,1}$, and $[\Sigma_R]_{NN}\!=\!A_{N,N+1}~g_{R}~A_{N+1,N}$.
Here, $A_{0 1}$ [$A_{N+1,N}$] represents the coupling Hamiltonian between the last [first] supercell of the left [right] contact and the first [last] supercell of the channel. 
%--------------------------------
The supercell arrangement for FV-NEGF has two other major advantages as: (I) it allows the implementation of the Sancho-Rubio method~\cite{sancho1985highly} to speed up the evaluation of $g_L$ and $g_R$, (II) it \textcolor{black}{enables employing} the direct recursive algorithm~\cite{tho2014recur,nguyen2023recursive}, which should be implemented to make \textcolor{black}{
full 3D quantum transport simulations computationally practical.}
%---------------------
If we define $[G_C]\!=\![\overline{G}_C][\Delta V_C]$, the final matrix form of the retarded Green's function becomes identical to the conventional FD-NEGF or TB-NEGF formulation. While this may appear trivial, the construction of $\left[A_C\!-\!\Sigma_L \!-\!\Sigma_R\right]$ by the FV method is non-trivial due to the enforcement of conservation laws at all local mesh interfaces. 
%---------------------
The remainder of the quantum transport theory---including the evaluation of observables such as the transmission function (terminal current) and local density of states (charge distribution) in terms of the retarded $G^R$, advanced $G^A$, and lesser ($G^<$) Green's functions---remains unchanged. We therefore omit further theoretical details here, as they can be found elsewhere \cite{thakur2023tutorial}.

%%%%%%%%%%%%%%%%%%%%%%%%%%%%%%%%%%%%%%%%%%%%%%%%
\section{\label{sec:level3} Representative 
Applications for FV-NEGF}
\begin{figure}[h]
	\begin{center}
	\includegraphics[width=8.0cm]{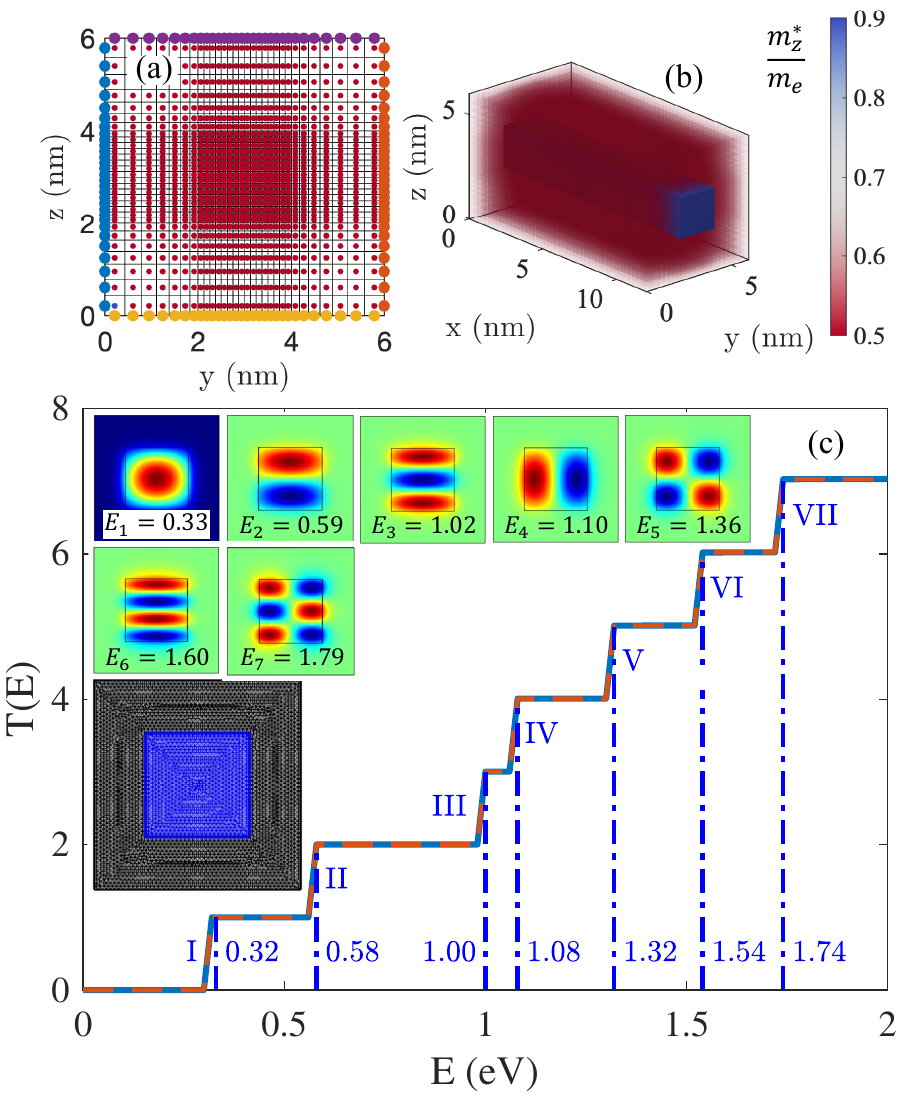}
	\end{center}
	\caption{\label{fig3} (a) Cross-sectional FV mesh. The dots on the circumference are boundary dots. (b) The ratio $m_z^*/m_e$ in 3D as an example for non-isotropic space-dependent material properties. (c) Transmission function evaluated by FV-NEGF. FE method in 2D is used to calculate cross sectional modes shown as inset color plots. FE mesh is also plotted as an inset.}
\end{figure}

 To verify the capability of our FV-NEGF method, we consider electron transport through a 3D core-clad nanowire. The nanowire consists of a silicon (Si) core with a cross-sectional area of \textcolor{black}{$2\!\times\!2~nm^2$} in the $yz$-plane, embedded in a silicon dioxide (SiO$_2$) cladding with a \textcolor{black}{$2~nm$} thickness. Here, the wire extends from $0$ to \textcolor{black}{$12~nm$} along the $x$-axis. The [$100$] crystal orientation of Si is aligned along the $z$-axis. Thus, the effective masses in the Si core are defined as $m_z^*\!=\!0.9$ (longitudinal) and $m_{x,y}^*\!=\!0.2$ (transverse). For the SiO$_2$ cladding, we set $m_{x,y,z}^*\!=\!0.5$ (isotropic). The conduction band offset between Si and SiO$_2$, $U$, is set to 3.1 $eV$. In Figs.~\ref{fig3}(a) and (b), we show our cross-sectional FV mesh and $m_z^*$ distribution in the 3D domain.  
Here, the number of 2D cells in the $yz$-plane is 900, and along the x-axis there are 40 grid spaces. 
%--------------------
\textcolor{black}{Quantum transport study is then carried out by a home-built recursive NEGF code.}
%--------------------
\textcolor{black}{In Fig.~\ref{fig3}(c), we plot the transmission function, $T(E)$,} for a limited energy range such that seven steps are included. We remind readers that the current-voltage characteristic of a quantum wire at low temperature follows $T(E)$ which shows the typical step-like increase (conductance step) as the electron energy $(E)$ (bias voltage) increases. Each conductance step corresponds to the involvement of a new cross-sectional mode in the transport. Transport modes can be evaluated by solving the Schr\"odinger equation with the same input parameters in the 2D cross-sectional domain (closed system). In addition, seven cross-sectional modes and corresponding eigenenergies are calculated with the finite-element method and plotted as insets in Fig.~\ref{fig3}(c). 
%------------------
It is clear that the onset of the steps aligns with the eigenenergies denoted by $E_i$, thereby confirming the correctness of the methodology and its numerical implementation.
%%%-------
The LDOS(E) distribution in the cross-sectional $yz$-plane resembles the mode distributions shown in Fig.~\ref{fig3}(c) (data not displayed for brevity). 
%(a) Cross-sectional FV mesh. The dots on the circumference are boundary dots. 
\begin{figure}[h]
	\begin{center}
		\includegraphics[width=8.0cm]{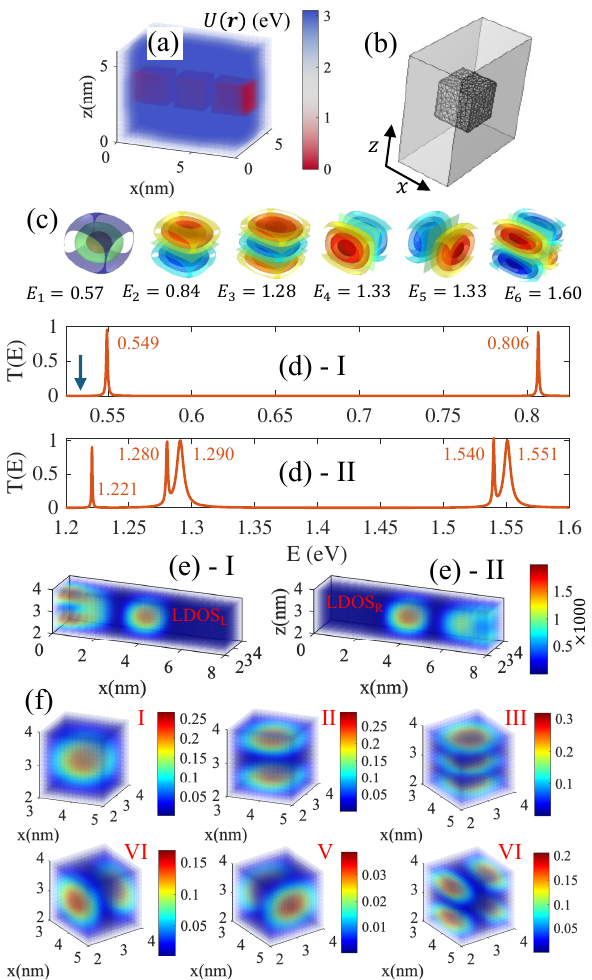}
	\end{center}
	\caption{\label{fig4} (a) $U(\mathbf{r})$ (electron affinity) in 3D domain. (b) FE mesh in the 0DEG domain. (c) Six bound states and their corresponding eigenenergies calculated by FE method. (d) Transmission function calculated by FV-NEGF. (e)-I [II] LDOS from left (LDOS$_L$) [right (LDOS$_R$)] to right [left] at an energy denoted by an arrow in (d). (f) LDOS$_L$ at six resonance peaks in (d).}
\end{figure}
%The integral of LDOS(E) over energy, $\int \text{LDOS(E)}dE$, gives rise to the electron density (1DEG). 
Here, we refrain from quantifying the electron density, as our analysis does not involve sweeping the source-drain electrochemical potentials.
%%%%%%%%%%%%%%%%%%%%%%%%%%%%%%%%%%%%%%%%%%%%%%
%---------------------------------------------------------------------------------
For the second example, we introduce two $0.5~nm$ \textcolor{black}{thick} SiO$_2$ separator layers (barrier) in the $yz$-plane. In this configuration, the left and right 1D wires (each $2~nm$ long) connect to a central $2~nm$ isolated quantum dot (0DEG), forming a 1DEG-0DEG-1DEG system, \textcolor{black}{ as shown in} Fig.~\ref{fig4}(a).
Here, we modify the meshing style along the x-axis, such that grid space becomes much finer ($0.125~nm$) in the area between the two barriers. 
%--------------------
The transport characteristics of this system can be understood as follows:  
(I) \textcolor{black}{electrons tunnel from the cross-sectional modes of the left 1DEG into the bound states of the 0DEG (a fully confined region), then into} the right 1DEG (or vice versa), following parallel tunneling mechanism.  
(II) \textcolor{black}{peaks in T(E) are expected near the 0DEG eigenenergies, but their broadening and energy shifts depend on the barrier properties between the 1DEG wires and the 0DEG (dot), following resonant tunneling mechanism.} 
%--------------------
Here, due to the high potential barrier, the broadening can be very small, requiring a fine energy grid ($dE$) to resolve the resonant peaks accurately. \textcolor{black}{To address this,} we have used the FE method in 3D and numerically compute the first six bound states of the 0DEG, \textcolor{black}{as shown in} Fig.~\ref{fig4} (b) and (c). We then adopt much finer $dE$s near these energies. In fact, we intentionally choose thin separators ($0.5~nm$) to induce slight broadening in the resonance peaks. 
%-------------
\textcolor{black}{In Fig.~\ref{fig4}(d) (I) and (II), resonance peaks in $T(E)$ are depicted with high resolution. The resonance energy shifts can be obtained by subtracting the eigenenergies shown in Fig.~\ref{fig4}(c) from the resonance energies presented in Fig.~\ref{fig4}(d) (I) and (II). These shifts are on the order of a few tens of meV. In addition, doubly degenerate levels (at $E_4\!=\!E_5$ and $E_6\!=\!E_7$) split when the 0DEG is connected to the 1DEGs on the right and left.} 
%----------------
The \textcolor{black}{local density of state (LDOS) presents the other key result}. 
%----------------
\textcolor{black}{In Fig.~\ref{fig4}(e) (I) and (II), we plot the LDOS from the left and right contacts (LDOS$_L$ and LDOS$_R$) at an energy just below the first resonance peak [the arrow in Fig.~\ref{fig4}(d)-I], revealing coupling between mode 2 of the 1DEGs and the first bound state of 0DEG. These 3D plots indicate how the leakage of LDOS from the right and left contacts contributes to the total electron density given by $n(\mathbf{r})\!=\!\sum_{\alpha\in L,R}\int \operatorname{LDOS}_{\alpha} f(\mu_{\alpha}) (dE/2\pi)$, where $f$ is the fermi function and $\mu$ is the electrochemical potential associated with contacts.}
%%%%%%%%%%%%
Fig.~\ref{fig4}(f) displays LDOS$_L$ at the resonance peaks in the 0DEG area. The 3D distributions match the FE-calculated bound states (Fig.~\ref{fig4}(c)), but with the presence of energy shifts. The broadening effect becomes more pronounced [peak 5 in Fig.~\ref{fig4}(d)] when the bound states misaligned with the cross-sectional modes in the 1DEG \textcolor{black}{(Fig.~\ref{fig4}(f)-V)}. 
%%%%%%%%%%%%%%%%%%%%%%%%%%%%%%%%%%%%%%%%%%%%%%%%%%%%%%%
%%%%%%%%%%%%%%%%%%%%%%%%%%%%%%%%%%%%%%%%%%%%%%%%%%%%%%%
\section{\label{sec:level4} Conclusion}
In summary, we have proposed a cell-centered Finite-Volume implementation of the NEGF approach (FV-NEGF) for modeling quantum transport in low-dimensional devices. The most significant advantage of the FV-NEGF method \textcolor{black}{lies in} its exceptional simplicity when applied to quantum transport problems in 3D domains. Our approach inherently accommodates nonuniform meshes, making it particularly suitable for mesoscopic systems. 
%-----------------
Furthermore, we establish a connection between the mesoscopic FV-NEGF framework and the microscopic tight-binding NEGF (TB-NEGF) method. This new implementation is particularly promising for modeling disordered systems, as it incorporates material properties into quantum transport by assigning material constants to the centers of FV cells and strictly enforcing local conservation laws.
%------------------------
We validate the FV-NEGF method through two representative examples that would be challenging to implement using conventional FD/FE-NEGF methods. 
%------------------------
Although we did not explore fully coupled self-consistent Poisson-NEGF simulations in this work, we are confident that the FV-NEGF method would perform successfully in such scenarios. The possible self-consistent approach enables the use of a unified mesh for both the NEGF and Poisson equations, making consistent numerical treatment. 
%%%%%%%%%%%%%%%%%%%%%%%%%%%%%%%%%%%%%%%%%%%%%%%%%%%

\section*{Conflict of interest}
There are no conflicts to declare.

\section*{Data availability statement}
The data that support the findings of this study are available
upon reasonable request from the authors.

\section*{Acknowledgements}
\label{sec5}
We thank Wei Zhu and Kun Yang for very useful conversations. V.M acknowledges funding from Summer Academy Program for International Young Scientists (Grant No. GZWZ[2022]019). W.D acknowledges the startup funding from Westlake University. 
%%%%%%%%%%%%%%%%%%%%%%%%%%%%%%%%%%%%%%%%%%%%AAAAAAAAAAAAAAAA
%----------------------------------------------------------------
%%%%%%%%%%%%%%%%%%%%%%%%
% \newpage
\section*{References}
\bibliographystyle{iopart-num}
\bibliography{bib_fv_negf.bib}
%%%%%%%%%%%%%%%%%%%%%%%%
\end{document}